\documentclass[aps,preprint,preprintnumbers,amsmath,amssymb,nofootinbib]{revtex4}

\usepackage{amssymb}

\usepackage{lscape}

\usepackage{amsmath}

\usepackage{bm}

\usepackage{graphicx}

\usepackage{epsfig}

\usepackage{color}

\begin{document}

\title{Exponential collapse with variable time scale driven by a scalar field}

\author{$^{1}$ Jaime M. Hern\'andez\footnote{E-mail: jaime.mendoza@alumno.udg.mx}, $^{2,3}$ Mauricio Bellini
\footnote{{\bf Corresponding author}: mbellini@mdp.edu.ar}, $^{1}$ Claudia Moreno\footnote{E-mail: claudia.moreno@cucei.udg.mx} }
\address{$^1$ Departamento de F\'{\i}sica, Centro Universitario de Ciencias Exactas e Ingenier\'{\i}as, Universidad de Guadalajara
Av. Revoluci\'on 1500, Colonia Ol\'impica C.P. 44430, Guadalajara, Jalisco, M\'exico. \\
$^2$ Departamento de F\'{\i}sica, Facultad de Ciencias Exactas y
Naturales, Universidad Nacional de Mar del Plata, Funes 3350, C.P. 7600, Mar del Plata, Argentina.\\
$^3$ Instituto de Investigaciones F\'{\i}sicas de Mar del Plata (IFIMAR), \\
Consejo Nacional de Investigaciones Cient\'ificas y T\'ecnicas
(CONICET), Mar del Plata, Argentina.}

\begin{abstract}
We study the dynamic collapse driven by a scalar field, when a relativistic observer falls co-moving with the collapse and cross the horizon of a Schwarzschild black-hole (BH), at $t=t_0$. During the collapse the scale of time is considered as variable. Back-reaction effects and gravitational waves produced during the exponential collapse are studied. We demonstrate that back-reaction effects act as the source of gravitational waves emitted during the collapse, and wavelengths of gravitational waves (GW) are in the range: $\lambda \ll r_s\equiv {e^{-2h_0t_0}\over 2 h_0}$, that is, smaller than the Schwarzschild radius. We demonstrate that during all the collapse the global topology of the space-time remains hyperbolic when the observer cross the horizon.
\end{abstract}

\maketitle

\section{Introduction and motivation}

One of the reasons to study extended proposals of General Relativity (GR) is the conceptual need to incorporate the quantum limit of gravity into the theory \cite{qg}. The second one is to describe the origin of the cosmological (variable) constant in the universe, which has decreased $183$ orders of magnitude since the big bang \cite{b1}. In particular, the consequences of non-trivial space time topology for the laws of physics has been a topic of great interest for theoretical physicists \cite{weyl}. Very interesting proposals were made in order to incorporate spinor fields in a framework of a quantum field theory \cite{ber,ber1,ber2}. In a more recent proposal, non-trivial spatial topology has been introduced in order to describe the space time in a quantum framework, which can be accorded with an extended General Relativity theory in an unified description of quantum and classical interactions in physics with spin incorporated \cite{uni}, called Unified Spinor Fields (USF).

The understanding of a collapsing system is a very important issue in theoretical physics, mainly in astrophysics and cosmology physics. Some extensive investigations were made related to a spherically symmetric collapse driven by a scalar field \cite{Gundlach}, and also taking into account quantum gravitational effects \cite{HW}. More recently, this issue has been analytically explored in \cite{GJ,G}. On the other hand, a study of a collapse of a fluid with a heat flux has been treated in \cite{Sharma} and the dynamics of dissipative gravitational collapse has been studied in\cite{HS}. An interesting issue to study is the evolution of the global topology of space time during a collapse driven by a scalar field that can avoid the final singularity \cite{B,J}. This topic was explored in a recent work \cite{col} jointly with the
geometrical back-reaction of space time produced by this collapse. During the collapse, the system reaches a tiny equation of state $\omega=1$ with a spectral index: $n_s= 0$, which favors the detection on wavelengths with size $\lambda \ll a_0$, smaller than the initial size of the collapsing ball from a point of view of a co-moving observer. Of course, the model would be more realistic if the collapse were considered as inhomogeneous. This case with heat conduction with a generalized scalar field distribution was explored recently in some works \cite{c1,c2}, but without consider a variable timescale. In this work we are aimed to study a collapsing system in which can be
possible to match exterior and interior solutions, by considering a complex time $d\tau = e^{i\theta(t)} dt$, to describe the topological evolution of space time, when the observer cross the horizon, at $t=t_0$. In this framework, the proposal consists in evaluate if the space time remains hyperbolic during the collapse.

This work is organized as follow: in Sect. II we revisit Relativistic Quantum Geometry (RQG). In Sect. III we develop the example of exponential collapse with variable timescale, where we study the classical dynamics, as well as the back-reaction effects, joined with the emission of gravitational waves during the collapse. Finally, in Sect. IV we develop some final comments about the results.

\section{Relativistic Quantum Geometry revisited}

Following some recent work, in this one we shall consider an extended General Relativity proposal, where the theory is constructed on a extended manifold such that, in the sense of the Riemannian geometry, the covariant derivative of the metric tensor is null, so that $\Delta g_{\alpha\beta}=g_{\alpha\beta;\gamma} \,dx^{\gamma}=0$
(we denote with $;$ the Riemann-covariant derivative). The extended manifold is obtained by making a displacement from the background Riemann manifold to the manifold described
by the connections
\begin{equation}\label{ga}
\Gamma^{\alpha}_{\beta\gamma} = \left\{ \begin{array}{cc}  \alpha \, \\ \beta \, \gamma  \end{array} \right\}+ g_{\beta\gamma} \sigma^{\alpha},
\end{equation}
on which the non-metricity is nonzero, so that
\begin{equation}\label{gab}
\delta g_{\alpha\beta} = g_{\alpha\beta|\gamma} \,dx^{\gamma} = -\left[\sigma_{\beta} g_{\alpha\gamma} +\sigma_{\alpha} g_{\beta\gamma}
\right]\,dx^{\gamma},
\end{equation}
where $dx^{\alpha}$ is the eigenvalue that results when we apply the operator $ \delta\check{x}^{\alpha} (x^{\beta}) $ on a background quantum state $ \left. | B \right> $\cite{rb}, defined on the Riemannian manifold:
\begin{equation}
dx^{\alpha} \left. | B \right> =  \bar{U}^{\alpha} dS \left. | B \right>= \delta\check{x}^{\alpha} (x^{\beta}) \left. | B \right> .
\end{equation}
The quantum state defined on the Riemann background is represented in an ordinary Fock space. We denote with a {\it  bar} quantities represented on the Riemann background manifold. Furthermore, the scalar field $\sigma(x^{\alpha})$ drives a geometrical displacement from a Riemann manifold to an extended one, such that the action
\begin{equation}\label{ac}
{\cal I} = \int d^4 x\, \sqrt{-\bar{g}}\, \left[\frac{\bar{R}}{2\kappa} + \bar{{\cal L}}\right] = \int d^4 x\, \left[\sqrt{-\bar{g}} e^{-2\sigma}\right]\,
\left\{\left[\frac{\bar{R}}{2\kappa} + \bar{{\cal L}}\right]\,e^{2\sigma}\right\},
\end{equation}
remains invariant. Hence, if we require that $\delta {\cal I} =0$, we obtain
\begin{equation}
-\frac{\delta V}{V} = \frac{\delta \left[\frac{\bar{R}}{2\kappa} + \bar{{\cal L}}\right]}{\left[\frac{\bar{R}}{2\kappa} + \bar{{\cal L}}\right]}
= 2 \,\delta\sigma,
\end{equation}
where $\delta\sigma = \sigma_{\mu} dx^{\mu}$ is an exact differential and $V=\sqrt{-\bar{ g}}$ is the volume of the Riemann manifold. All the variations are defined on the extended manifold, and gauge invariance is assured because $\delta {\cal I} =0$. The perturbed metric with
back-reaction effects included is
\begin{equation}\label{met1}
g_{\mu\nu} = {\rm diag}\left[\bar{g}_{00}\, e^{2\sigma}, \bar{g}_{11}\, e^{-2\sigma}, \bar{g}_{22}\, e^{-2\sigma},\bar{g}_{33}\, e^{-2\sigma}\right],
\end{equation}
which preserves the invariance of the action.

\subsection{Relativistic dynamics on the extended manifold, quantum action and dynamics of $\sigma$}

The Einstein tensor can be written as \cite{rb}
\begin{equation}
{G}_{\alpha\beta} = \bar{G}_{\alpha\beta} + \sigma_{\alpha ; \beta} + \sigma_{\alpha} \sigma_{\beta} + \frac{1}{2} \,\bar{g}_{\alpha\beta}
\left[ \left(\sigma^{\mu}\right)_{;\mu} + \sigma_{\mu} \sigma^{\mu} \right],
\end{equation}
where we have made use of the fact that the connections are symmetric. Since the extended Einstein tensor is ${G}_{\alpha\beta} = \bar{G}_{\alpha\beta}- \Lambda(\sigma, \sigma_{\alpha})\, \bar{g}_{\alpha\beta}$, hence $\Lambda\equiv \Lambda(\sigma, \sigma_{\alpha})$ can be considered a functional on the extended manifold\cite{rb}
\begin{equation}\label{la}
\Lambda(\sigma, \sigma_{\alpha}) = -\frac{3}{4} \left[ \sigma_{\alpha} \sigma^{\alpha} + {\Box} \sigma\right],
\end{equation}
and we can define a geometrical quantum action on the extended manifold with the Lagrangian (\ref{la})
\begin{equation}
{\cal W} = \int d^4 x \, \sqrt{-g} \,\, \Lambda(\sigma, \sigma_{\alpha}),
\end{equation}
such that the dynamics of the geometrical field is given by the Euler-Lagrange equations, after imposing $\delta
\mathcal{W}=0$:
\begin{equation}
\frac{\delta \Lambda}{\delta \sigma} - \bar{\nabla}_{\alpha} \left( \frac{\delta \Lambda}{\delta \sigma_{\alpha}}\right) =0,
\end{equation}
where the variations are defined on the extended manifold. This means that $\delta\Lambda\neq 0$, but $\Delta\Lambda=0$. Furthermore, $ \Pi^{\alpha}=\frac{\delta \Lambda}{\delta \sigma_{\alpha}}=-{3\over 4} \sigma^{\alpha}$ is the geometrical momentum and the
dynamics of $\sigma$ describes a free scalar field
\begin{equation}\label{si}
{\Box} \sigma =0,
\end{equation}
so that the momentum components $\Pi^{\alpha}$ comply with
the equation
\begin{equation}
{\nabla}_{\alpha} \Pi^{\alpha} =0.
\end{equation}
If we define the scalar invariant
$\Pi^2=\Pi_{\alpha}\Pi^{\alpha}$, we obtain that
\begin{equation}
\left[\sigma,\Pi^{2}\right] =0,
\end{equation}
where we have used that $\sigma_{\alpha} U^{\alpha} = U_{\alpha} \sigma^{\alpha}$, and therefore one possible solution is
\begin{equation}\label{con}
\left[\sigma(x),\sigma^{\alpha}(y) \right] =- i \Theta^{\alpha}\, \delta^{(4)} (x-y), \qquad \left[\sigma(x),\sigma_{\alpha}(y) \right] =
i \Theta_{\alpha}\, \delta^{(4)} (x-y),
\end{equation}
with $\Theta^{\alpha} = \hbar\, \bar{U}^{\alpha}$ and $\Theta^2 = \Theta_{\alpha}
\Theta^{\alpha} = \hbar^2 \bar{U}_{\alpha}\, \bar{U}^{\alpha}$. Here, $\bar{U}^{\alpha}$ are the components of the Riemannian
velocities. Additionally, it is possible to define the Hamiltonian operator
\begin{equation}
{\cal H} = \left(\frac{\delta \Lambda}{\delta \sigma_{\alpha}}\right) \sigma_{\alpha} - \Lambda(\sigma,\sigma_{\alpha}),
\end{equation}
such that the eigenvalues of "quantum energy" becomes from ${\cal H}\left|B\right> = E\left|B\right>$. Can be demonstrated that
$\delta{\cal H}=0$, so that the quantum energy $E$ is an invariant on the extended manifold.

\subsection{Gravitational waves}\label{gravw}

In \cite{rb}, was proposed the existence of a tensor
field $\delta\Psi_{\alpha\beta}$, such that $\delta
R_{\alpha\beta}\equiv \nabla_{\beta} \delta W_{\alpha}-\delta\Phi
\,g_{\alpha\beta} \equiv \Box \delta\Psi_{\alpha\beta} -\delta\Phi
\,g_{\alpha\beta} =- \kappa \,\delta S_{\alpha\beta}$, and
hence $\delta W^{\alpha} = g^{\beta\gamma} \nabla^{\alpha}
\delta\Psi_{\beta\gamma}$, with $\nabla^{\alpha}
\delta\Psi_{\beta\gamma}=\delta\Gamma^{\alpha}_{\beta\gamma} -
\delta^{\alpha}_{\gamma} \delta\Gamma^{\epsilon}_{\beta\epsilon}$.
{\em Notice that the fields $\tilde{ \delta W}_{\alpha}$ and
$\tilde{\delta\Psi}_{\alpha\beta}$ are gauge-invariant under
transformations}:
\begin{equation}
\tilde{\delta W}_{\alpha} = \delta W_{\alpha} - \nabla_{\alpha} \delta\Phi, \qquad
\tilde{\delta\Psi}_{\alpha\beta} =\delta\Psi_{\alpha\beta} - \delta\Phi \,
g_{\alpha\beta}, \label{gauge}
\end{equation}
where $\delta\Phi=-\left(\Lambda/2\right)\,\delta\sigma$ complies $\Box \delta\Phi =0$. This means that exists a family of vector and tensor fields described by (\ref{gauge}), that are related to the Einstein tensor transformations
\begin{equation}
{G}_{\alpha\beta} = \bar{G}_{\alpha\beta} - \Lambda\, g_{\alpha\beta},
\end{equation}
and leave invariant the action (\ref{ac}). The gravitational wave equations are
\begin{equation}
\Box \tilde{\delta\Psi}_{\alpha\beta} =- \kappa \,\delta L_{\alpha\beta},
\end{equation}
with $L_{\alpha\beta} = 2 \frac{\delta {\cal \bar{L}}}{\delta g^{\alpha\beta}}$. The scalar field: $\tilde{\delta\Psi}=g^{\alpha\beta} \tilde{\delta\Psi}_{\alpha\beta}$, related to the gravitational waves components $\tilde{\delta\Psi}_{\alpha\beta} =\delta\Psi_{\alpha\beta} - \delta\Phi \,g_{\alpha\beta}$, is $\tilde{\delta\Psi} = \delta\Psi - 4 \delta\Phi$, such that the equation of motion for $\chi\equiv \frac{\tilde{\delta\Psi}}{\delta S}$, results to be
\begin{equation}\label{gw}
\Box \chi =- 2 \kappa\,\sigma^{\mu} U^{\beta}\, L_{\mu\beta}.
\end{equation}
In the case of a co-moving observer we must require $\bar{U}^i=0$ and $\bar{g}_{00}\,(\bar{U}^0)^2=1$.

\section{Collapse with variable timescale.}

We shall suppose that the timescale of the collapse is not constant, so that, a spatially flat, isotropic and homogeneous collapsing universe, can be represented by a line element

\begin{equation}\label{back}
d{S}^2 = e^{-2\int \Gamma(t)\, \,dt} \, e^{i\,{\pi\over 2} \left[1+{\bf Erf}\left(n(t-t_0)\right)\right] }\,dt^2 - \, e^{2\int h(t) dt}\,\, {\delta}_{ij}\, \,d{x}^i d{x}^j,
\end{equation}
that corresponds to a background metric that describes an observer which is falling with collapse and crosses the horizon of some black hole at $t=t_0$. Here, $h(t)<0$ is the collapse rate parameter on the background metric and $\Gamma(t)$ describes the timescale of the background metric, $n\gg 1$, is some constant sufficiently large, and ${\bf Erf}(x)={2\over \sqrt{\pi}} \int_{0}^{x}\,e^{-s^2}\,ds$ is the error function. In this paper, we shall consider natural units so that $c=\hbar=1$. In order to describe a collapse, we shall consider the action for a scalar field $\phi$ which is minimally coupled to gravity.

If its dynamics is governed by a scalar potential $V(\phi)$, the action can be written as

\begin{equation}\label{1}
{\cal I} = \int d^4x \, \sqrt{-\bar{g}} \,\left\{ \frac{{ \bar{R}}}{16\pi G} - \left[\frac{\dot\phi^2}{2}\,e^{2\int \Gamma(t)\,dt}  \, e^{-i\,{\pi\over 2} \left[1+{\bf Erf}\left(n(t-t_0)\right)\right] }\, - V(\phi)\right]\right\},
\end{equation}
where the volume of the background manifold is $\bar{v}=\sqrt{-\bar{g}}=\,e^{-\int \Gamma(t) \,dt}\,e^{3\int H(t)\, \,dt} \, e^{i\,{\pi\over 4} \left[1+{\bf Erf}\left(n(t-t_0)\right)\right] }\,$.

The action (\ref{1}) can be rewritten as

\begin{equation}\label{2}
{\cal I} = \int d^4x \, \sqrt{-\bar{g}}\,e^{2\int \Gamma(t)\,dt} \, e^{-i\,{\pi\over 2} \left[1+{\bf Erf}\left(n(t-t_0)\right)\right] }\,\,\left\{ \frac{ \bar{R}}{16\pi G} - \left[\frac{\dot\phi^2}{2} - \bar{V}(\phi)\right]\right\},
\end{equation}
that can be considered as an action for a minimally coupled to gravity scalar field on a background volume $\sqrt{-\bar{g}}\,e^{2\int \Gamma(t)\,dt} \, e^{-i\,{\pi\over 2} \left[1+{\bf Erf}\left(n(t-t_0)\right)\right] }\,$, a redefined potential $\bar{V}(\phi)=V(\phi)\,e^{-2\int \Gamma(t)\,dt} \, e^{i\,{\pi\over 2} \left[1+{\bf Erf}\left(n(t-t_0)\right)\right] }\,$, and an effective scalar curvature $\bar{R}= R\,e^{-2\int \Gamma(t)\,dt }\, e^{i\,{\pi\over 2} \left[1+{\bf Erf}\left(n(t-t_0)\right)\right] }\,$.

\subsection{background dynamics}

The effective volume of the background manifold in (\ref{2}), is $\bar{\bar{v}}=\sqrt{-\bar{g}} \,e^{2\int \Gamma(t)\,dt}\, e^{-i\,{\pi\over 2} \left[1+{\bf Erf}\left(n(t-t_0)\right)\right] }\,=e^{\int \Gamma(t)\, \,dt}\,e^{3\int h(t)\, \,dt}\, e^{-i\,{\pi\over 4} \left[1+{\bf Erf}\left(n(t-t_0)\right)\right] }\,$. The dynamics of the scalar field $\phi$, can be written in a compact form as
\begin{equation}
\ddot\phi + \left[3 h(t)+ \bar{\Gamma}(t)\right] \dot\phi +  \,\frac{\delta \bar{V}}{\delta\phi} =0. \label{infl}
\end{equation}

The background Einstein equations, are
\begin{eqnarray}
3 h^2\, e^{2\int  \bar{\Gamma}(t)\, \,dt}  &=& 8\pi\, G \,\rho, \label{a} \\
-\,\left[3 h^2 + 2 \dot{h} + 2 \bar{\Gamma}\, h\right] \, e^{2\int  \bar{\Gamma}(t)\, \,dt} &=& 8 \pi \,G\,P, \label{b}
\end{eqnarray}
where we have called $ \bar{\Gamma} = \Gamma - i\, \frac{\sqrt{\pi}}{2} n e^{-n^2(t-t_0)^2}$, $P= \left(\frac{\dot{\phi}^2}{2}-\bar{V}(\phi)\right)\, e^{2\int  \bar{\Gamma}(t)\, \,dt} $ is the pressure and $\rho=\left(\frac{\dot{\phi}^2}{2} + \bar{V}(\phi)\right)\, e^{2\int  \bar{\Gamma}(t)\, \,dt}$ the energy density due to the scalar field. Notice that $ \int  \bar{\Gamma}(t)\, \,dt = \int \Gamma(t)\,dt\,-i\,{\pi\over 4} \left[1+{\bf Erf}\left(n(t-t_0)\right)\right]$. Therefore, the system (\ref{a}) and (\ref{b}), hold
\begin{eqnarray}
3 h^2 &=& 8\pi\, G \,\left(\frac{\dot{\phi}^2}{2}+\bar{V}(\phi)\right), \label{a1} \\
-\left[3 h^2 + 2 \dot{h} + 2 \bar{\Gamma}\, h\right] &=& 8 \pi \,G\,\left(\frac{\dot{\phi}^2}{2}-\bar{V}(\phi)\right). \label{b1}
\end{eqnarray}

The equation of state that describes the dynamics of the system is:
\begin{equation}
\omega= \frac{P}{\rho} = -\left(1+\frac{2 \dot{h}}{3 h^2}+\frac{2 \bar{\Gamma}}{3 h}\right). \label{om}
\end{equation}

From the physical point of view, if we consider a co-moving frame where $U_{0}=\pm \sqrt{g_{00}}$ and $U_j=0$, such that $j$ can take the values $j=1,2,3$, the relativistic velocity, $U^0={dx^0\over dS}$, will describe the rate of time suffered by a relativistic observer which is falling with the collapse of the system, for an observer which is in a non-inertial frame. Notice that the velocities always will meet the normalization expression: $g_{\mu\nu}\, U^{\mu}\,U^{\nu}=1$. After working with Eqs. (\ref{a1}) and (\ref{b1}), we obtain
\begin{eqnarray}
\dot\phi &=& \pm \sqrt{-\frac{1}{4\pi G} \left( \dot{h} + h \bar{\Gamma}\right)}, \label{a2} \\
\bar{V} & = & \frac{1}{8\pi G} \left( 3 h^2 + \dot{h} + h \bar{\Gamma} \right). \label{b2}
\end{eqnarray}

\subsection{An example: Exponential collapse}

As an example, we consider the case where the Hubble parameter, which is negative during the collapse, is a constant:
\begin{equation}
h(t)= -h_0.
\end{equation}
In this case the general solution for ${\Gamma}(t)$, which comply with the equation (\ref{infl}), is
\begin{eqnarray}
{\Gamma}(t) &=& \frac{i\,n}{2{\sqrt {\pi }}}\,\left(\frac{C\,\cosh{\left(\frac{3h_0}{2}t\right)}
+\sinh{\left(\frac{3h_0}{2}t\right)}}{C\,\cosh{\left(\frac{3h_0}{2}t\right)}+\sinh{\left(\frac{3h_0}{2}t\right)}}\right) {\rm e}^{-\left[{n}^{2}(t-t_0)^{2} \right]} \nonumber \\
&+& \frac{3}{2} h_0\left[C+1\right]\left(\frac{\cosh{\left(\frac{3h_0}{2}t\right)}
+\sinh{\left(\frac{3h_0}{2}t\right)}}{C\,\cosh{\left(\frac{3h_0}{2}t\right)}+\sinh{\left(\frac{3h_0}{2}t\right)}}\right).
\end{eqnarray}
If we take the particular solution $C=1$, we obtain that
\begin{equation}
{\Gamma}(t) = \frac{i\,n}{2{\sqrt {\pi }}}\,  {\rm e}^{-\left[{n}^{2}(t-t_0)^{2} \right]}+3\,h_0,
\end{equation}
and hence $\overline{\Gamma}(t)=3\,h_0$. Therefore, during the collapse the global topology of the space-time remains hyperbolic when the observer cross the horizon at $t=t_0$, and causality is preserved for the falling observer. The equation of state for this model of collapse is constant
\begin{equation}
\omega(t) =1 .
\end{equation}


The effective potential, $\bar{V}$, is null
\begin{equation}
\bar{V}(t)= 0,\label{pot}
\end{equation}
choosing the positive root in (\ref{a2}), we obtain
\begin{equation}
{\dot \phi}(t)= \left[\frac{3 h^2_0}{4 \pi G} \right]^{\frac{1}{2}}.
\end{equation}

The dynamics of the back-reaction effects is described by the equation
\begin{equation}
\ddot{\sigma}+\left(\overline{\Gamma}-3h_{0}\right)\dot{\sigma}-e^{2\int (h_{0}-{\Gamma})dt }\,e^{i\,{\pi\over 2} \left[1+{\bf Erf}\left[n(t-t_0)\right]\right]}\,\nabla^{2}\sigma=0.
\end{equation}

The geometrical scalar field $\sigma$ can be expressed in terms of the fourier expansion
\begin{equation}
\sigma(x^{i},t)=\frac{1}{(2\pi)^{3/2}}\int d^{3}k \left[A_{k}e^{i\vec{k}\cdot \vec{x}}\xi_{k}(t)+A_{k}^{\dagger}e^{-i \vec{k}\cdot \vec{x}}\xi_{k}^{*}(t)\right],
\end{equation}
with creation and annihilation operators $A^{\dagger}_k$ and $A_k$, that comply with the quantum algebra
\begin{equation}\label{m5}
\left<B\left|\left[A_{k},A_{k'}^{\dagger}\right]\right|B\right>=\delta ^{(3)}(\vec{k}-\vec{k'}),\quad
\left<B\left|\left[A_{k},A_{k'}\right]\right|B\right>=\left<B\left|\left[A_{k}^{\dagger},A_{k'}^{\dagger}\right]\right|B\right>=0.
\end{equation}
When timescale is variable, the metric with back-reaction effects included results to be
\begin{equation}\label{met1}
g_{\mu\nu} = {\rm diag}\left[e^{-2\int \bar{\Gamma}(t)dt}\, e^{2\sigma}, -\, e^{-2h_0\,t}\, e^{-2\sigma}, - \, e^{-2h_0\,t}\, e^{-2\sigma}, - \, e^{-2h_0\,t}\, e^{-2\sigma}\right].
\end{equation}
The equation of motion of the modes $\xi$ is
\begin{equation}
\ddot{\xi}_{k}+\left(\overline{\Gamma}-3h_0\right)\dot{\xi}_{k}+k^{2} e^{2\int (h_{0}-{\Gamma})dt }\,e^{i\,{\pi\over 2} \left[1+{\bf Erf}\left[n(t-t_0)\right]\right]} \xi_{k}=0,
\end{equation}
which results to be
\begin{equation}
\ddot{\xi}_{k}(t)+k^2\,e^{-4h_0 t}\,{\xi}_{k}(t)=0. \label{dif}
\end{equation}
Notice that the integrate of $\Gamma(t)$, is
\begin{equation}
\int_{t_0}^{t} \Gamma(t) = i\,{\pi\over 4} \left[1+{\bf Erf}\left[n(t-t_0)\right]\right]+3\,h_0\,\left(t-t_0\right).
\end{equation}
The general solution of the Eq. (\ref{dif}), is
\begin{equation}
\xi_k(t)= A\, {\cal Y}_0\left[x(t)\right] + B\,{\cal Y}_0\left[x(t)\right] ,
\end{equation}
where $x(t)= { N}\,e^{-2h_0 (t-t_0)}$, for ${k\over 2h_0} = N\,e^{2h_0\,t_0}$. Here, $N$ is a dimensionless parameter which give us the
wave-number $k$ re-scaled as a multiple of the Schwarzschild radius, which is crossed by the mode at $t=t_0$. In order to the modes $\xi_k(t)$ to
be well defined for all $t$, we must require that $B=0$.
Also must know the normalization condition for the modes $\xi_{k}(t)$. The nonzero component of the relativistic velocity is $\bar{U}^{0}=\frac{dx^{0}}{dS}$, and, because we use $c=1$, we obtain that $\frac{dx^{0}}{d\tau}=\frac{1}{\sqrt{g_{00}}}=\sqrt{g^{00}}$, so that $\bar{U}^{0}=\sqrt{g^{00}}$, and therefore the quantization of modes $\xi(t)$ is
\begin{equation}
\dot{\xi}^{*}_{k}(t)\xi_{t}(t)-\xi^{*}_{k}(t)\dot{\xi}_{k}(t)=i\, U^{0}=i\sqrt{g^{00}}=ie^{\int\overline{\Gamma}(t)dt}.
\end{equation}
In our example, we have $\overline{\Gamma}(t)=3h_{0}$, so that we have
\begin{equation}
\dot{\xi}^{*}_{k}(t)\xi_{t}(t)-\xi^{*}_{k}(t)\dot{\xi}_{k}(t)=ie^{3h_{0}t}.
\end{equation}
The solution of Eq. (\ref{dif}), once quantized, is\footnote{In order to quantize, we use the asymptotic expression $\mathcal{Y}_{0}[x(t)]\simeq -i\frac{1}{\sqrt{2\pi x(t)}}e^{-i(x(t)-\pi/4)}$, which is valid when $x(t)\gg1$.}
\begin{equation}
\xi_{k}(t)=i\sqrt{\frac{\pi}{2h_{0}}}e^{\frac{3}{2}h_{0}t}\mathcal{Y}_{0}\left[Ne^{-2h_{0}(t-t_{0})}\right].
\end{equation}
The time derivative of the time dependent modes, $\xi(t)$, is
\begin{equation}
\dot{\xi}_{k}(t)=i\sqrt{\frac{\pi}{2h_{0}}}e^{\frac{3}{2}h_{0}t}\left[\frac{3}{2}h_{0}\,t\,\mathcal{Y}_{0}
\left[x(t)\right]+2N\,h_{0}e^{-2h_{0}(t-t_{0})}\mathcal{Y}_{1}\left[x(t)\right]\right],
\end{equation}
and therefore, the squared norm related to $\dot{\xi}_{k}(t)$, is
\begin{eqnarray}
\left|\dot{\xi}_{k}(t)\right|^2 &=& \pi h_{0} e^{3h_{0}t}\left[(9/8)t^{2}\mathcal{Y}^{2}_{0}\left[x(t)\right]+3N\,t\,e^{-2h_{0}(t-t_{0})}\mathcal{Y}_{0}\left[x(t)\right]
\mathcal{Y}_{1}\left[x(t)\right]\right.\nonumber \\
&+& \left.2N^{2}e^{-4h_{0}(t-t_{0})}\mathcal{Y}^{2}_{1}\left[x(t)\right]\right].
\end{eqnarray}
Hence, the power spectrum of the energy density fluctuations due to back reaction effects, is
\begin{equation}\label{power}
\mathcal{P}_{\dot\sigma}=\frac{k^{3}h_{0}e^{3h_{0}t}}{2\pi}\left[(9/8)t^{2}\mathcal{Y}^{2}_{0}\left[x(t)\right]
+3N\,t\,e^{-2h_{0}(t-t_{0})}\mathcal{Y}_{0}\left[x(t)\right]\mathcal{Y}_{1}\left[x(t)\right]
+2N^{2}e^{-4h_{0}(t-t_{0})}\mathcal{Y}^{2}_{1}\left[x(t)\right]\right],
\end{equation}
where the last expression is valid in all scales.

\subsection{Large scales power spectrum}

In order to study the large scales power spectrum, we must calculate the asymptotic limit of the Bessel functions in the expression (\ref{power}) when $x(t)\ll 1$, which are
\begin{equation}\label{1aprox}
\mathcal{Y}_{0}\left[x(t)\right]^{2}\simeq\frac{4}{\pi^{2}}\left[\ln\left(\frac{x(t)}{2}\right)+\gamma\right]^{2},
\end{equation}
\begin{equation}\label{2aprox}
\mathcal{Y}_{0}\left[x(t)\right]\mathcal{Y}_{1}\left[x(t)\right]\simeq-\frac{4\Gamma(1)}{\pi^{2} x(t)}\left[\ln\left(\frac{x(t)}{2}\right)+\gamma\right],
\end{equation}
\begin{equation}\label{3aprox}
\mathcal{Y}^{2}_{1}\left[x(t)\right]\simeq\frac{4\Gamma^{2}(1)}{\pi^{2}x^{2}(t)}.
\end{equation}
Here, $\gamma$ is the Euler-Mascheroni constant. When the $x(t)\ll 1$, the terms (\ref{1aprox}) and (\ref{3aprox}) becomes dominant and (\ref{2aprox}) goes to zero, so dropping out (\ref{2aprox}),  the large scales power spectrum of the energy density fluctuations due to back-reaction effects is\footnote{Here, we use $Y_{0}(x(t))\simeq \frac{2}{\pi}\left[ln(x(t)/2)+\gamma\right]$ and $Y_{1}(x(t))\simeq-\frac{\Gamma(1)}{\pi}(2/x(t))$.}
\begin{equation}
\left.\mathcal{P}_{\dot\sigma}\right|_{x(t)\ll1}=\frac{4k^{3}h_{0}
e^{3h_{0}t}}{\pi^{3}}\left[\left(\ln\left(\frac{N}{2}\right)-2h_{0}(t-t_{0})+\gamma\right)^{2}+\Gamma^{2}(1)\right].
\end{equation}
The amplitude of these power-spectrum increases with time.

\subsection{Small scales power spectrum}

In order to study the small scales power spectrum, we must calculate the asymptotic limit of the Bessel functions in the expression (\ref{power}) when $x(t)\gg 1$, obtaining
\begin{equation}\label{4aprox}
\mathcal{Y}^{2}_{0}\left[x(t)\right]\simeq\frac{2}{\pi x(t)}\sin^{2}\left(x(t)-\frac{\pi}{4}\right),
\end{equation}
\begin{equation}\label{5aprox}
\mathcal{Y}_{0}\left[x(t)\right]\mathcal{Y}_{1}\left[x(t)\right]\simeq\frac{2}{\pi x(t)}\sin\left(x(t)-\frac{\pi}{4}\right)\sin\left(x(t)-\frac{3\pi}{4}\right),
\end{equation}
\begin{equation}\label{6aprox}
\mathcal{Y}^{2}_{1}\left[x(t)\right]\simeq\frac{2}{\pi x(t)}\sin^{2}\left(x(t)-\frac{3\pi}{4}\right),
\end{equation}
so the power spectrum at small scales is
\begin{eqnarray}
\left.\mathcal{P}_{\dot\sigma}\right|_{x(t)\gg 1}&=& \frac{k^{3}h_{0}}{N\pi^{2}}e^{3h_{0}t}e^{2h_0(t-t0)}  \left[(9/8) t^{2}\sin^{2}(x(t)-\pi/4)\right.\nonumber \\
&+& 3\,N\,t\,e^{-2h_{0}(t-t_{0})}\sin(x(t)-\pi/4)\sin(x(t)-3\pi/4) \nonumber \\
&+& \left.2N^{2}
e^{-4h_{0}(t-t_{0})}\sin^{2}(x(t)-3\pi/4)\right].
\end{eqnarray}
Notice that, as in the large-scale power spectrum, the amplitude increases exponentially with time.

\section{Variation of the energy density fluctuations}

As was calculated in previous works \cite{rb}, the variation of the energy density fluctuations is
\begin{equation}\label{fluc}
\left\langle B\left\vert \frac{1}{\overline{\rho}}\frac{\delta\overline{\rho}}{\delta S}\right\vert B\right\rangle=-2U^{0}\dot{\sigma},
\end{equation}
where $\dot{\sigma}=\left\langle \dot{\sigma}^{2} \right\rangle^{1/2}$, and the square fluctuations $\left\langle \dot{\sigma}^{2} \right\rangle$ are given by
\begin{equation}
\left\langle \dot{\sigma}^{2} \right\rangle \simeq \int_{0}^{\epsilon k_{0}t}\frac{dk}{k}\mathcal{P}_{\dot{\sigma}}.
\end{equation}
In order to study the variation of the fluctuations at both scales (i.e. large and small scales), we will use approximations at both scales to calculate the square fluctuations.

\subsection{Variation of the energy density fluctuations at large scales}

To calculate those fluctuations at large scales, we shall use the approximation of Eq.(\ref{power}) when $x(t)\ll 1$ into the formula
\begin{equation}
\left\langle \dot{\sigma}^{2} \right\rangle_{x(t)\ll 1} \simeq \int_{0}^{\epsilon k_{0}t}\frac{dk}{k}\mathcal{P}_{\dot{\sigma}},
\end{equation}
obtaining
\begin{equation}
\left\langle \dot{\sigma}^{2} \right\rangle_{x(t)\ll 1} =\frac{4\epsilon^{3}k_{0}^{3}t^{3}h_{0}e^{3h_{0}t}}{3\pi^{3}}\left[\left(\ln\left(\frac{N}{2}\right)
-2h_{0}(t-t_{0})+\gamma\right)^{2}+\Gamma^{2}(1)\right].
\end{equation}
Therefore, taking the squared root in the last equation and plugging into Eq. (\ref{fluc}), we obtain
\begin{equation}
\left< B\left| \frac{1}{\overline{\rho}}\frac{\delta\overline{\rho}}{\delta S}\right|B\right>=-2\sqrt{\frac{4\epsilon^{3}k_{0}^{3}t^{3}h_{0}}{3\pi^{3}}}e^{(9/2)\,h_{0}t}
\left[\left(\ln\left(\frac{N}{2}\right)-2h_{0}(t-t_{0})+\gamma\right)^{2}+\Gamma^{2}(1)\right]^{1/2}.
\end{equation}

\subsection{Variation of the energy density fluctuations at small scales}

To calculate those fluctuations at large scales we shall use the approximation of the Ec. (\ref{power}) when $x(t)\gg 1$ into the formula
\begin{equation}
\left\langle \dot{\sigma}^{2} \right\rangle_{x(t)\gg1} \simeq \int_{0}^{\epsilon k_{0}t}\frac{dk}{k}\mathcal{P}_{\dot{\sigma}},
\end{equation}
obtaining
\begin{eqnarray}
\left\langle \dot{\sigma}^{2} \right\rangle_{x(t)\gg1}&=& \frac{\epsilon^{3}k^{3}_{0}t^{3}h_{0}}{3\pi^{2}N}e^{3h_{0}t}e^{2h_{0}(t-t_{0})}\left[(9/8) t^{2}\,\sin^{2}(x(t)-\pi/4)\right. \nonumber \\
&+& 3N\,te^{-2h_{0}(t-t_{0})}\sin(x(t)-\pi/4)\sin(x(t)-3\pi/4)    \nonumber \\
&+& \left. 2N^{2}e^{-4h_{0}(t-t_{0})}\sin^{2}(x(t)-3\pi/4)\right].
\end{eqnarray}
Hence, taking the squared root in the last equation and plugging into the Eq. (\ref{fluc}), we obtain
\begin{eqnarray}
\left< B\left| \frac{1}{\overline{\rho}}\frac{\delta\overline{\rho}}{\delta S}\right| B\right> \, & = & - 2\sqrt{\frac{\epsilon^{3}k^{3}_{0}t^{3}h_{0}}{3\pi^{2}N}}\,e^{\frac{9}{2} h_{0}t} e^{h_{0}(t-t_{0})} \left[(9/8) t^{2}\,\sin^{2}(x(t)-\pi/4)
\right. \nonumber \\
&+& \left. 3N\,t\,e^{-2h_{0}(t-t_{0})}\sin{(x(t)-\pi/4)}\sin(x(t)-3\pi/4)\right. \nonumber \\
& +& \left. 2 N^{2}\,e^{-4h_{0}(t-t_{0})}\sin^{2}{(x(t)-3\pi/4)}\right]^{1/2}.
\end{eqnarray}
Notice that in both cases, small and large scales, the energy density fluctuations increase with time with the collapse, so that the system becomes more and more unstable. These instabilities are the seed for the emission of gravitational waves, which we shall study in the following subsection.

\subsection{Gravitational waves from the exponential collapse}

In order to describe GW during the exponential collapse, we shall use the formalism revisited in (\ref{gravw}), with the equation (\ref{gw}), which in our example holds:
\begin{equation}\label{es}
\ddot{\chi} -e^{-4h_0\,t} \nabla^2\chi =2\kappa\,\dot\sigma \bar{U}^0 \left(\dot\phi\right)^2 ,
\end{equation}
where we have used that $\Box \chi$ is
\begin{equation}
\Box\chi \equiv \bar{g}^{\alpha\beta} \nabla_{\alpha}\, \chi_{,\beta}=\bar{g}^{\alpha\beta} \left[ \chi_{,\alpha\beta} -  \left\{ \begin{array}{cc}  \epsilon \, \\ \alpha \, \beta  \end{array} \right\} \chi_{,\epsilon}\right].
\end{equation}
The general solution for the modes $\chi_k(\vec{x},t) = B_k\,e^{i\vec{k}.\vec{x}}\,f_k(t)$, for $B_k=A_k$, is
\begin{eqnarray}
f_k(t) & = & C_1\,{\cal J}_0[x(t)]+C_2\,{\cal Y}_0[x(t)] \nonumber \\
&-& \frac{3}{2} i\,\frac{\left( h_0 \pi\right)^{3/2}{\cal J}_0[x(t)]}{\sqrt{2}} \left\{  \int dt \left\{e^{(9/2) h_0 t} {\cal J}_0[x(t)]
\, \left[ 4Ne^{-2h_0(t-t_0)} {\cal Y}_1[x(t)] + 3t\, {\cal Y}_0[x(t)]\right]\right\} \right.\nonumber \\
&+&\left. \int dt\, \left\{ e^{(9/2) h_0 t} {\cal Y}_0[x(t)]
 \, \left[ 4Ne^{-2h_0(t-t_0)} {\cal Y}_1[x(t)] + 3t\, {\cal Y}_0[x(t)]\right]\right\}\right\},  \label{os}
\end{eqnarray}
where $C_1$ and $C_2$ are constants.

\begin{figure}
  \centering
    \includegraphics[width=13cm, height=14cm]{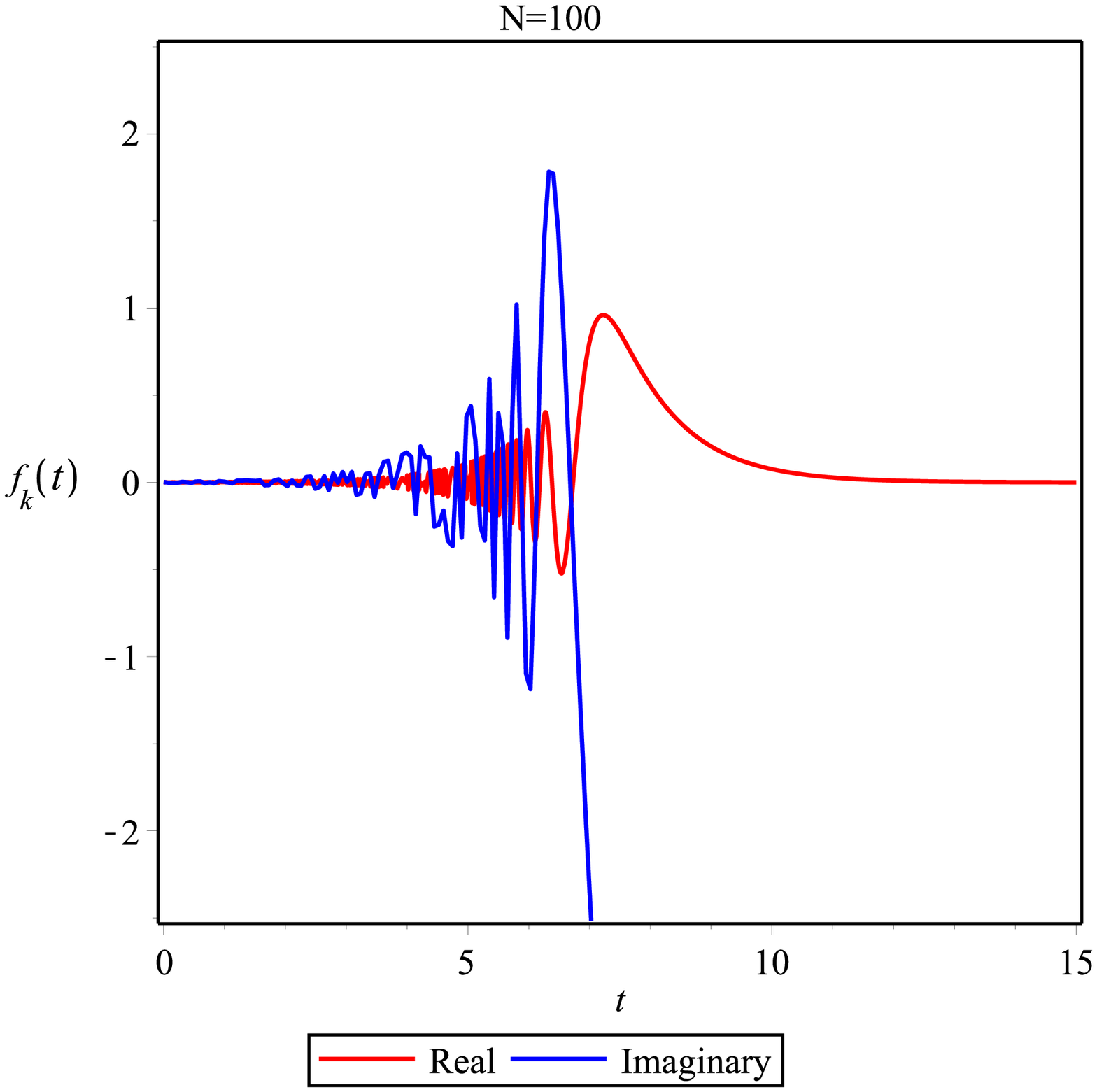}
  \caption{Plotting of the real (red line) and imaginary (blue line) parts of $f_k(t)$ for $N=100$ and $t_{0}=5$. Notice that for $t>t_0$ the real contribution goes to zero, but the imaginary part of $f_k(t)$ tends to $-\infty$.}
  \label{f1}
\end{figure}

\begin{figure}
  \centering
    \includegraphics[width=13cm, height=14cm]{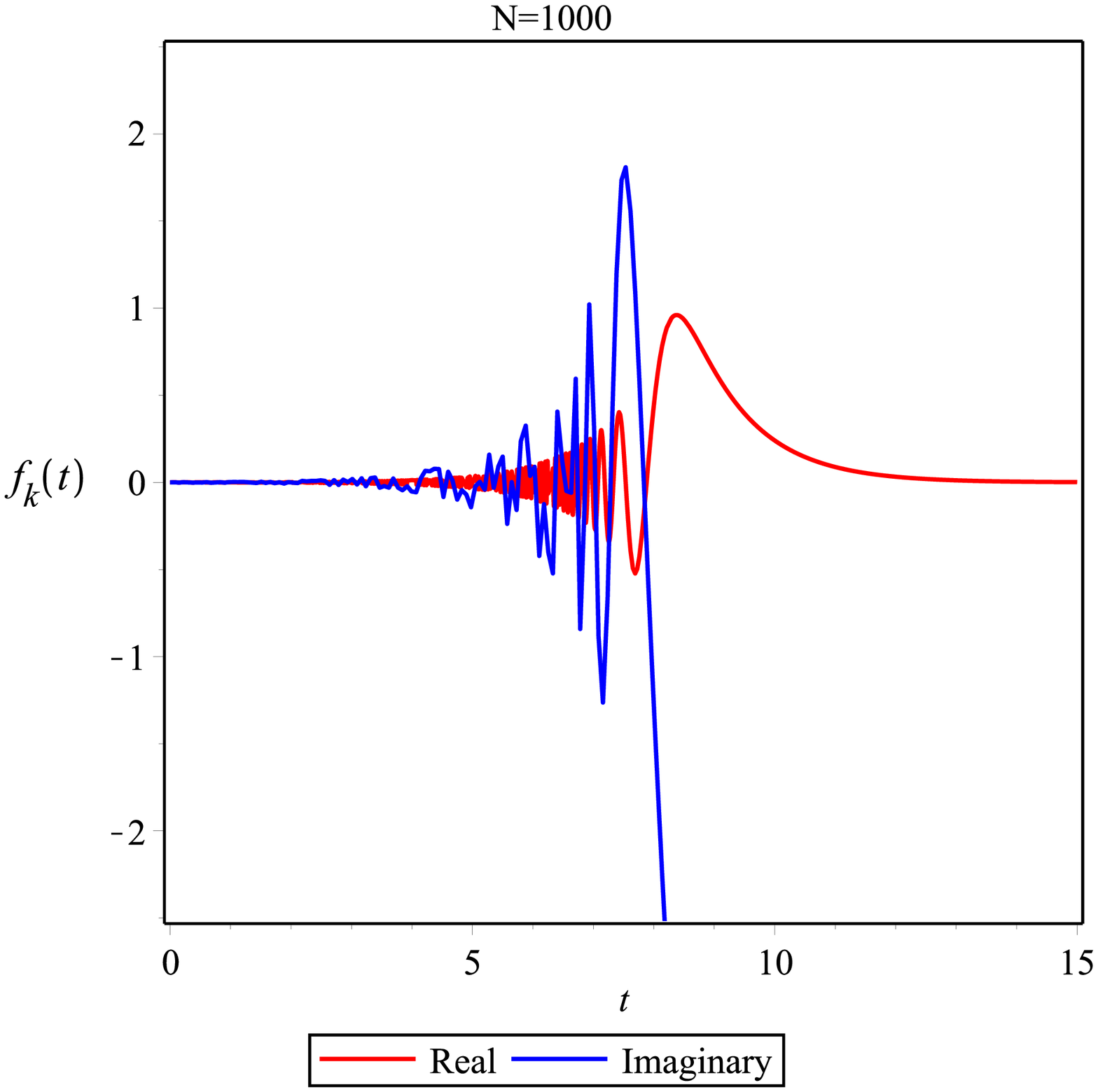}
  \caption{Plotting of the real (red line) and imaginary (blue line) parts of $f_k(t)$ for $N=1000$ and $t_{0}=5$. Notice that for $t>t_0$ the real contribution goes to zero, but the imaginary part of $f_k(t)$ tends to $-\infty$. The real part of $f_k$ decays later than for $N=100$.}
  \label{f2}
\end{figure}

If we consider that sources emit signals with wavelengths smaller than the size of the black-hole, the inhomogeneous part of the differential equation (\ref{es}), must be considered with $x\gg 1$ in the solution. Hence, we can use asymptotic forms of the Bessel functions to solve the integrals in (\ref{os}). Using $\mathcal{J}_{\alpha}[x(t)]\approx\sqrt{\frac{2}{\pi x(t)}}\cos[x(t)-\alpha\pi/2-\pi/4]$ and $\mathcal{Y}_{\alpha}[x(t)]\approx\sqrt{\frac{2}{\pi x(t)}}\sin[x(t)-\alpha\pi/2-\pi/4]$, after expanding sines and cosines into a power series, we obtain
\begin{eqnarray}\label{ogs2}
f_k(t)&=&\sqrt{\frac{2}{\pi C_{x}}}e^{h_{0}t}\left[C_1\cos(C_{x}e^{-2h_{0}t}-\frac{\pi}{4})+C_2\sin(C_{x}e^{-2h_{0}t}-\frac{\pi}{4})\right] \nonumber \\
&-&\frac{3}{2} i\,\frac{\left( h_0 \pi\right)^{3/2}}{\sqrt{\pi C_{x}}}e^{h_{0}t}\cos(C_{x}e^{-2h_{0}t}-\pi/4)\left[\mathcal{I}_{1}+\mathcal{I}_{2}+\mathcal{I}_{3}+\mathcal{I}_{4}\right],
\end{eqnarray}
being
\begin{eqnarray}
\mathcal{I}_{1}&=&-\frac{4}{\pi}\int e^{\frac{9}{2}h_{0}t}\left[\sin\left(2C_{x}e^{-2h_{0}t}\right)+1\right]\,dt , \label{I1}\\
\mathcal{I}_{2}&=&-\frac{3}{N\pi}e^{-2h_{0}t_{0}}\int te^{\frac{13}{2}h_{0}t}\left[\cos\left(2C_{x}e^{-2h_{0}t}\right)\right]\,dt , \label{I2} \\
\mathcal{I}_{3}&=&\frac{4}{\pi}\int e^{\frac{9}{2}h_{0}t}\cos\left((2C_{x}e^{-2h_{0}t}\right)\,dt , \label{I3} \\
\mathcal{I}_{4}&=&\frac{3}{2N\pi}e^{-2h_{0}t_{0}}\int t\,e^{\frac{13}{2}h_{0}t}\left[1-\sin\left((2C_{x})e^{-2h_{0}t}\right)\right]\,dt. \label{I4} \\
\end{eqnarray}
Here, $C_{x}=N\,e^{2h_{0}t_{0}}$ is a dimensionless constant. In order to solve the equations (\ref{I1})-(\ref{I4}), we can expand the sines and cosines in power series:
\begin{eqnarray}
\mathcal{I}_{1}&=&-\frac{4}{\pi}\left[\Sigma_{n=0}^{\infty}\frac{(-1)^{n}(2C_{x})^{2n+1}}{(2n+1)!}\frac{e^{\left[\left(\frac{9}{2}
-(2h_{0})(2n+1)\right)h_{0}t\right]}}{h_{0}\left[\frac{9}{2}-2(2n+1)\right]}+\frac{e^{\frac{9}{2}h_{0}t}}{\frac{9}{2}h_{0}}\right], \nonumber \\
\mathcal{I}_{2}&=&-\frac{3}{N\pi}e^{-2h_{0}t_{0}}\left[\Sigma_{n=0}^{\infty}\frac{(-1)^{n}(2C_{x})^{2n}}{(2n)!}
\frac{e^{\left[\left(\frac{13}{2}-4n\right)h_{0} t\right]}(\frac{13}{2}h_{0}t-4h_{0}n-1)}{h^2_{0}(\frac{13}{2}-4n)^{2}}\right], \nonumber \\
\mathcal{I}_{3}&=&\frac{4}{\pi}\left[\Sigma_{n=0}^{\infty}\frac{(-1)^{n}(2C_{x})^{2n}}{(2n)!}
\frac{e^{\left[\left(\frac{9}{2}-4n\,\right)h_{0} t\right]}}{h_{0}\left(\frac{9}{2}-4n\right)}\right], \nonumber \\
\mathcal{I}_{4}&=&\frac{3}{2N\pi}e^{-2h_{0}t_{0}}\left[\frac{4e^{\frac{13}{2}h_{0}t}(\frac{13}{2}h_{0}t-1)}{(13h_{0})^{2}}\right. \nonumber \\
&-& \left.\Sigma_{n=0}^{\infty}\frac{(-1)^{n}(2C_{x})^{2n+1}}{(2n+1)!}\frac{e^{\left[\left(\frac{13}{2}
-2(2n+1)\right)h_{0} t\right]}\left[\frac{13}{2}h_{0}t-2h_{0}(2n+1)-1\right]}{h^2_{0}\left[\frac{13}{2}-2(2n+1)\right]^{2}}\right].
\end{eqnarray}
In the figures (\ref{f1}) and (\ref{f2}) we have plotted the imaginary and real contributions of the modes $f_k(t)$ for $N=100$ and $N=1000$, that correspond to wavelengths $\lambda \equiv {\pi e^{-2h_0t_0}\over h_0\,N} \ll {e^{-2h_0t_0}\over 2 h_0}\equiv r_s$, which are smaller than the Schwarzschild radius, $r_s$. Notice that in both cases the real contribution become zero after cross the horizon, but the imaginary contribution increases with time. This is because the source becomes more intense with time as the observer cross the Schwarzschild radius. However, the real part of $f_k$ decays later for largest wavenumber $N$.

\section{Final Comments}

We have studied an exponential collapse driven by a scalar field for a co-moving relativistic observer which is falling with the collapse. In order to study the evolution of the global space-time topology, we have proposed a variable timescale in which the time rotates on a complex plane during the collapse. During the collapse the equation of state remains $\omega=-1$. The interesting is that during all the collapse the global topology of the space-time remains hyperbolic when the observer cross the horizon at $t=t_0$, so that causality is preserved for the falling relativistic observer. With respect to the timescale, in the example we have studied the system collapse with a effective background metric
\begin{displaymath}
dS^2 = e^{-6h_0(t-t_0)}\,dt^2 - e^{-2h_0\,t} \,\delta_{ij}\,dx^i\,dx^j,
\end{displaymath}
and hence the effective time scale is governed by the expression: $d\tau= \sqrt{\bar{g}_{00}}\, dt=e^{-3h_0(t-t_0)}\,dt$. Therefore, for
$t<t_0$ the time run accelerated, but for $t>t_0$ (i.e., after the observer cross the Schwarzschild horizon), the physical time $\tau$ decelerates and the observer never reaches the center of the black-hole. This would be seen by a distant observer as the accumulation of dark matter in the interior of the black-hole. In other words, black-holes can be considered as dark matter.

Finally, we have studied GW produced during the implosion, taking into account back-reaction effects during the exponential collapse. For the scalar description of these waves [see eq. (\ref{gw})]: $\Box \chi =- 2 \kappa\,\sigma^{\mu} \bar{U}^{\beta}\, L_{\mu\beta}$, we have demonstrated that back-reaction effects (through the field $\sigma^{\alpha}$), are the source of GW with wavelengths of the order of (or smaller), the horizon scale. This implies that the relevant wave numbers will be $k\gg 2 h_0\,N\,e^{2h_0t_0}$, that corresponds with wavelengths produced by the black-hole that is in the range: $\lambda \ll r_s\equiv {e^{-2h_0t_0}\over 2 h_0}$, that is, smaller than the BH.

\section*{Acknowledgements}

\noindent M. B. acknowledges CONICET, Argentina (PIP 11220150100072CO) and UNMdP (EXA852/18) for financial support. This research was supported by the CONACyT-UDG Network Project No. 294625 "Agujeros Negros y Ondas Gravitatorias".

\end{document}